\documentclass[aps,prb,twocolumn,groupedaddress,showpacs]{revtex4}

\usepackage{graphicx}
\usepackage{amsmath}

\newcommand {\eq}{%
  \begin{equation}}
\newcommand {\qe}{%
  \end{equation}}
\newcommand {\eqa}{%
  \begin{eqnarray}}
\newcommand {\aqe}{%
  \end{eqnarray}}

\begin{document}

\title{Probability of the resistive state formation caused by absorption of a single-photon in current-carrying superconducting nano-strips}

\author{Alexei Semenov}
\author{Andreas Engel}
\altaffiliation{Physics Institute of the University of Zurich, Winterthurerstr. 190, 5087 Zurich, Switzerland}
\author{Heinz-Wilhelm\ H\"{u}bers}
\affiliation{DLR Institute of Planetary Research, Rutherfordstr.\ 2, 12489 Berlin, Germany}

\author{Konstantin Il'in}
\author{Michael Siegel}
\affiliation{Institute of Micro- and Nano-Electronic Systems, University of Karlsruhe, Hertzstr.\ 16, 76187 Karlsruhe, Germany}


\begin{abstract}
We have studied supercurrent-assisted formation of the
resistive state in nano-structured Nb and NbN superconducting
films after absorption of a single photon. In amorphous narrow NbN
strips the probability of the resistive state formation has a
pronounced spectral cut-off. The corresponding threshold photon energy
decreases with the bias current. Analysis of the experimental data
in the framework of the generalized hot-spot model suggests that
the quantum yield for near-infrared photons increases faster than
the photon energy. Relaxation of the resistive state depends on
the photon energy making the phenomenon feasible for the
development of energy resolving single-photon detectors.
\end{abstract}


\maketitle

\section{Introduction}
\label{intro} A recently proposed hot-spot scenario of the
resistive state formation in a thin, narrow, current-carrying
superconducting strip due to absorption of a single infrared
photon \cite{Semenov1} was subsequently suggested as a mechanism
of single-photon detection \cite{Goltsman1,Verevkin1} in NbN
microbridges. Although soon after realization such detectors found
applications, the first experimental evidence of the envisaged
mechanism has been reported only a year ago \cite{Zhang1}. The
hot-spot model \cite{Semenov1} describes the formation and the
following evolution of a normal hot-spot around the absorption site in an
infinite homogeneous superconducting film where a single photon
has been absorbed. The energy of the captured photon is quickly
shared among a large number of electrons via electron-electron and electron-phonon
interaction leading to an avalanche-like multiplication
\cite{Kozorezov1} of non-equilibrium quasiparticles (QP). The
normal spot appears when the excess QP concentration becomes large
enough in order to locally destroy superconductivity. At some
stage the normal spot reaches the maximum size that is a trade-off
between quasiparticle multiplication and their diffusion out of
the absorption site. In a narrow strip carrying a small
sub-critical current, no resistive state is formed unless the
maximum size of the normal spot $A_n$ exceeds the strip width. If
the spot-size is less than the strip-width, the resistive state
may appear due to a supplementary action of the bias current that
should have a density $j$ close to the critical current density
$j_C$. Once the normal spot becomes larger than the coherence
length, the supercurrent is expelled from the spot. The effective
cross-section available for the current flow decreases and the
actual current density may exceed the critical current density.
When this happens, a normal domain spans the entire cross-section
of the strip that causes a voltage pulse developing between the
strip ends. The number of photons, which result in a voltage
response, related to the number of photons crossing the
geometrical area of the strip is a measure of the quantum
efficiency (probability) of the resistive state formation. If
$A_n$ is not sufficiently large in order to enhance the current
density to its critical value, there is still a finite probability
of the resistive state formation due to multi-photon or
fluctuation assisted events \cite{Goltsman1,Verevkin1}. Since
$A_n$ increases with the photon energy there should be a current
dependent photon energy $\epsilon_{0}$ that demarcates these two
regimes of the resistive state formation. The threshold energy
$\epsilon_{0}$ is defined by the simple geometric criteria
$A_n(\epsilon_{0})/w = 1-(j/j_C)$ where $w$ is the width of the
strip. For photon energies $\epsilon
> \epsilon_{0}$ but smaller than $h\nu_0$ ($\nu_0$ is the plasma frequency), the
quantum efficiency should equal the probability of photon
absorption and should not, therefore, depend on the photon energy.
At $\epsilon < \epsilon_{0}$ the quantum efficiency is dominated
by multi-photon and fluctuation assisted events, which generally
have a smaller probability.

So far the cut-off of the quantum efficiency at $\epsilon \approx
\epsilon_{0}$ has not been experimentally observed, instead a
monotonous decrease of the quantum efficiency with the photon
wavelength has been reported \cite{Verevkin1}. Knee-like features
have been found in the overall exponential dependence of the
quantum efficiency on the bias current. Although they were
attributed to the above discussed cut-off, the concluded size of
the normal spot was far too large in comparison to model estimates
\cite{Semenov2}. The photon absorption in a superconductor obeys
Poisson statistics, in that the probability of absorbing
simultaneously $n$ photons from a weak photon flux is proportional
to the flux intensity to the power of $n$. Variation of the
quantum efficiency corresponding to $n = 1, 2$ and 3 have indeed
been found \cite{Goltsman1} in NbN microbridges. However, this
observation would generally manifest quantum nature of any
detector and is not uniquely specific to any particular response
mechanism. A time delay of the resistive state formation in NbN
meander lines has been recently reported \cite{Zhang1}. The delay
was attributed to the supplementary action of the bias current in
the formation of the normal domain but the quantitative
description implied the initial normal spot much larger than the
hot-spot model predicts.

In this work we report measurements of the quantum efficiency of
the resistive state formation in narrow amorphous strips made from Nb
and NbN films. A reduced electron diffusivity and a reduced
density of electronic states in NbN allowed us to observe the
predicted cut-off of the quantum efficiency and to evaluate the
intrinsic efficiency of the electron multiplication. We refine the
hot-spot model and show that the resistive state formation in the
absence of fluctuations may be triggered by a single photon even
though no normal spot appears. This approach eliminates
contradictions within the earlier published experimental data and
provides a quantitative explanation of how fluctuations enhance
the quantum efficiency beyond the cut-off. We also present recent
results showing that the resistive state created by a photon
``remembers'' its energy, i.e. the effect has energy resolving
capability.

\section{Experiment}
\label{sec:1}
\subsection{Sample preparation and measurement technique}
\label{sec:2}
\begin{figure}
\begin{center}
\resizebox{0.4\textwidth}{!}{%
  \includegraphics{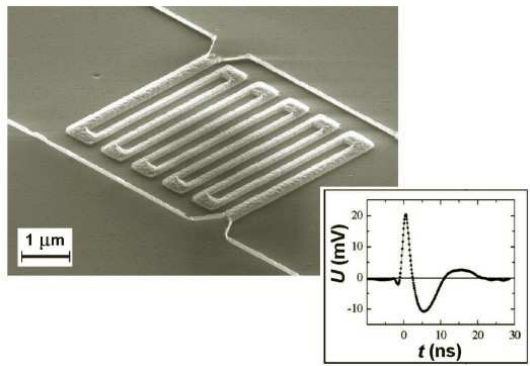}
}
\end{center}
\vspace{-0.4cm}       
\caption{SEM photo of a representative Nb meander on Si
substrate. Visible fringes at the edges of the structure are
formed by photoresist left for device protection. Inset shows a
typical voltage pulse as recorded by the oscilloscope.}
\label{fig:1}       
\end{figure}

Experimental results were obtained with narrow
meander lines made from either Nb or NbN thin films. Meanders
covering an area of 4 x 4 $\mu$m$^2$ were used in order to increase
optical coupling with near infrared photons. Contrary to the
conventional technique, our modified NbN films were prepared by dc
magnetron sputtering of an Nb target in an Ar+N$_2$ gas admixture
that had a reduced partial pressure of N$_2$. The deposition regime in combination with
relatively high sputtering rates of about 1.2 nm/sec resulted in
amorphous films with a nitrogen content smaller than in the
stoichiometric composition. Films with a nominal
thickness of 5  nm were deposited on sapphire substrates kept at
room temperature. Mostly because of the nitrogen
deficiency, the superconducting transition in 5 nm thick films
occurred at $\approx{6}$ K that is about one third of the
transition temperature of bulk NbN. Moreover, the transition
temperature was partly suppressed due to the proximity effect
\cite{Cooper1} between the central superconducting layer of the
film and both the non-superconducting oxidized surface on one side
and the film-substrate boundary on the other side. Films were
patterned using electron-beam lithography and ion milling. Our Nb
films had a thickness of 10 nm and were prepared on Si substrates
by dc magnetron sputtering in pure Ar atmosphere. Films were
patterned using electron-beam lithography and reactive ion
etching. The line width was 84 nm and 98 nm and the normal state
resistivity just above the superconducting transition was
$6\cdot10^{-6}\ \mathrm{\Omega}\,$cm and $6.7\cdot10^{-4}\ \mathrm{\Omega}\,$cm for Nb and
NbN meanders, respectively. Due to a partial damage of strip edges
by argon ions during the etching process, a portion of the meander
line near the edges became normal. These normal areas further
decreased \cite{Rem1} the transition temperature. For both
materials the transition temperature $T_C$ of the meander varied
between 4 and 5~K. An image of a representative Nb meander made
with a scanning electron microscope is shown in Fig.\ \ref{fig:1}.

In order to ensure a sufficiently high quality of our NbN
structures we evaluated their superconducting parameters relevant
to the formation of a critical current state. Measuring the
temperature variations of the second critical magnetic field
$B_{c2}$ near $T_C$, we concluded the diffusion coefficient $D =
0.35$ cm$^2$ sec$^{-1}$. This relatively low electron diffusivity
assured that the meanders were in the diffusive, dirty limit.
Extrapolating the linear temperature dependence of $B_{c2}$ near
$T_C$, we found $B_{c2}(0) = 9.4$ T and the zero-temperature
coherence length $\xi_0 = (\Phi_0/\pi B_{c2}(0))^{1/2}\approx8.5$
nm. The density of electronic states $N_0 = 2.2\cdot 10^{24}$
m$^{-3}$ K$^{-1}$ is then deduced from the Einstein's relation
$N_0 = 1/(e^2 \ \rho \ D)$ where $\rho$ stands for the
normal-state resistivity. Further assuming a classical energy gap
$\Delta_0 = 1.76\,k_B\,T_C$, we estimated the London penetration
depth $\lambda_L(0) = 1200$ nm. Since the density of states was an
important parameter for the evaluation of the size of the normal
spot, we cross-checked it measuring the critical current density
as a function of temperature. Because the width of our structure
was well below the London penetration depth, the supercurrent
density did not vary across the meander line. To derive the
critical current density from the measured values of the critical
current we used the electrical cross-section of the meander. Due
to the proximity effect at the lateral interfaces and the damaged
edges the electrical cross-section was almost one half \cite{Rem1}
of the geometrical cross-section. The best fit of the measured
temperature dependence with $j(T) = 3.27 e N_0 \Delta_0 (k_B T_C
D/h)^{1/2} (1-(T/T_C)^2)(1-(T/T_C)^4)^{1/2}$ resulted in $N_0 =
2.7\cdot 10^{24}$ m$^{-3}$ K$^{-1}$ in agreement with the density
of states inferred from the film resistivity. Practical
coincidence of the measured critical current with the depairing
critical current evidences amorphous structure with a large
concentration of defects that play a role of pinning centers. The
granularity, if any, does not influence the critical current and
should not therefore effect diffusion of electrons.

The substrate carrying the meander was thermally anchored to the
cold plate of a He$^4$--bath cryostat and was operated at
temperatures ranging from about 1.8~K to approximately 3.0~K. The
meander was illuminated using an incandescent light source. To
enable spectral measurements, the light was passed through a prism
monochromator. The meander was voltage-biased through a voltage
divider mounted inside the cryostat. Estimated time constant of
the bias was $\tau_B=0.5$~ns. The response to illumination was in the form
of a random sequence of voltage pulses each of them signaling
either absorption of a photon or a dark count event. Voltage
pulses were amplified using broadband microwave amplifiers (noise
temperature 6~K, band-pass from 0.1~GHz to 1.6~GHz) and then
guided either to a 250~MHz bandwidth/1~GHz sampling rate
oscilloscope or to a 200~MHz bandwidth voltage-level counter.
Inset in Fig.\ \ref{fig:1} shows the resulting signal transient recorded with
the oscilloscope. The mean photon count rate, that is proportional
to the quantum efficiency, was normalized to the light intensity
measured with either a silicon photodiode or a PbSe photodiode (at
longer wavelengths). For dark count measurements the optical
entrance to the cryostat was blocked completely.

Relaxation of the resistive state, and consequently the voltage
pulse across the meander, should last longer when initiated by a
photon with a larger energy. Material parameters of NbN, in
particular electron-phonon interaction time, suggest that at 2~K
the duration of the voltage pulse should be in the sub-nanosecond
range. Bearing in mind the band-pass of our read-out electronic,
we expect it to integrate pulses in real time. The amplitude
of the signal transient after the amplifier chain should then
correlate with the energy of the absorbed photon. Varying the
discriminator level of the counter, we measured the statistical
distribution of the signal amplitude and retrieved the mean value
and the dispersion.

\subsection{Experimental data}
\label{sec:3} 
\begin{figure}
\begin{center}
\resizebox{0.35\textwidth}{!}{%
  \includegraphics{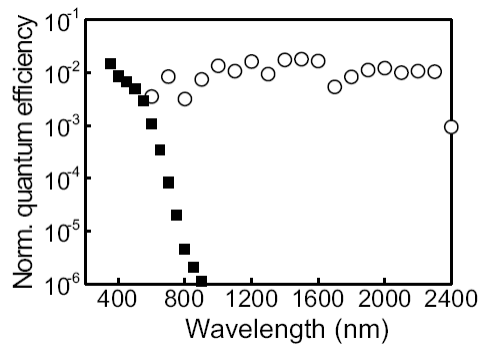}
}
\end{center}
\caption{Normalized quantum efficiency for Nb (closed symbols) and
NbN (open symbols) meanders.}
\label{fig:2}       
\end{figure}

The normalized quantum efficiencies of the resistive
state formation for Nb and NbN meanders are compared in Fig.\ \ref{fig:2}.
For both materials it was measured at 1.8~K using a current $I =
0.9 I_C$, where $I_C$ is the critical current at the operation
temperature. The quantum efficiency for the Nb meander drops
continuously with the increase of the wavelength showing no
threshold features while the NbN meander demonstrates an almost
constant efficiency of $\approx 1\%$ up to wavelengths that
exceed 2~$\mu$m. Obviously, the threshold photon energy for Nb
falls out of our experimental range and should correspond to a
wavelength less than 0.25~$\mu$m. According to the model described
in the next section, the threshold wavelength should be
proportional to $\rho(D)^{1/2}(\Delta_0^2\, d)^{-1}$. Given the
difference in the resistivity, the film thickness and the electron
diffusivity, more than an order of magnitude difference in the
threshold photon energy between Nb and NbN meanders is easily
understood. In the sections that follow we concentrate exclusively
on NbN meanders.

\begin{figure}
\begin{center}
\vspace{0.6cm}
\resizebox{0.35\textwidth}{!}{%
  \includegraphics{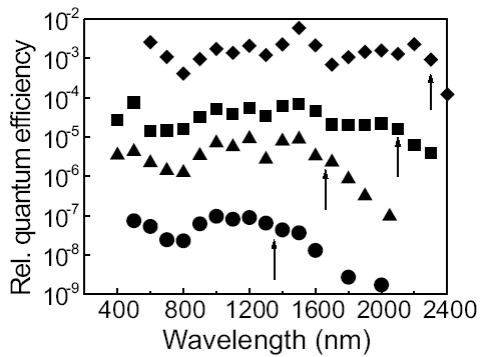}
}
\end{center}
\caption{Relative quantum efficiency for a representative NbN
meander biased with currents $0.6\,I_C$ , $0.77\,I_C$ , $0.8\,I_C$ and
$0.89\,I_C$. (from bottom to top). Arrows mark wavelengths
corresponding to the cut-off at each bias current. For
convenience, the data sets are shifted arbitrarily along the
vertical axis.}
\label{fig:3}       
\end{figure}

\begin{figure}
\begin{center}
\resizebox{0.38\textwidth}{!}{%
  \includegraphics{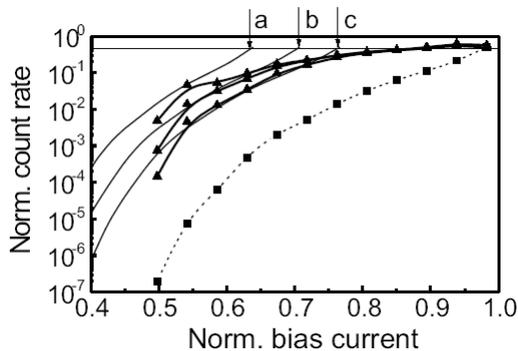}
}
\end{center}
\vspace{0.2cm}       
\caption{Normalized photon count rates for wavelengths 0.8~$\mu$m,
1.4~$\mu$m and 2~$\mu$m (thick solid lines, from top to bottom)
and dark count rate (dashed line) both measured for the NbN
meander at $\approx0.5\ T_C$. The largest absolute count rate was
$5\cdot10^7$~sec$^{-1}$. Arrows mark the currents corresponding to
the cut-off at: a -- 0.8~$\mu$m; b -- 1.4~$\mu$m; c -- 2~$\mu$m. Thin
solid lines approximate the photon count rates beyond the
cut-off.}
\label{fig:4}       
\end{figure}

A decrease of the current brings to light the cut-off of the
quantum efficiency in the NbN meanders. This can be seen in Fig.\ \ref{fig:3}
that shows the relative quantum efficiency of a representative NbN
meander measured at 2~K with different bias currents. For each
current there is a plateau in the wavelength dependence of the
quantum efficiency that is followed by a drop as the wavelength
increases. The wavelength corresponding to the cut-off increases
with the bias current. Alternatively, varying the current through
a meander exposed to illumination at a fixed wavelength, one
should also be able to observe the cut-off of the photon count
rate. However, this way additional uncertainties are introduced in detecting the threshold energy. The current destroys the
symmetry of the energy states of superconducting electron pairs.
The pair states enter the energy gap and the minimal excitation
energy for some electron pairs becomes smaller than the
zero-current energy gap. The excitation energy decreases with the
bias current and becomes zero when the current density reaches the
depairing critical current density. The decrease of the minimum
excitation energy with the bias current smears the threshold
feature and introduces a current dependence of the photon count
rate above the cut-off. Fig.\ \ref{fig:4} shows the current dependence of the
photon count rate (thick solid lines) for different photon
energies. Indeed, when the bias current is close to the critical
value, all curves merge and the count rate slowly varies with the
current. The curve corresponding to a particular wavelength starts
to deviate from this common dependence at a current that increases
with the wavelength. We associate the deviation from the common
dependence with the cut-off of the quantum efficiency. Threshold
values of the current are marked by arrows. The current dependence
of the photon count rate beyond the cut-off was approximated (thin
solid curves) using the current dependence of the dark
count rate shown as the dotted line in Fig.\ \ref{fig:4}. The
connection between the photon counts and dark counts will be
discussed below.

\section{Model refinement}
\label{sec:4} Applying to our experimental data (Fig.\ \ref{fig:3}) the
geometrical criteria of the hot-spot model, we find the maximum
normal-spot size of 8.5 nm created by a photon with the 1.2 $\mu$m
wavelength. An analytical approximation \cite{Semenov2} of the
numerical solution of the two dimensional diffusion problem gives
a close value of $\approx 9$ nm. According to this approximation,
even smaller normal spot should appear in conventional NbN films
\cite{Goltsman1,Verevkin1} because they have a larger density of
states, a larger energy gap and a larger electron diffusivity.
However, all those spots are comparable or less than the coherence
length suggesting that no resistive state should appear in
response to photons with larger wavelength. Contrary, our
experimental data show that not only the resistive state appear
but there is a well pronounced spectral cut-off corresponding to
wavelengths larger than 1.2 $\mu$m. In order to eliminate this
discrepancy, we will refine the model considering the effect of
quasiparticles, which are not confined in the normal spot.

\begin{figure}
\begin{center}
\resizebox{0.35\textwidth}{!}{%
  \includegraphics{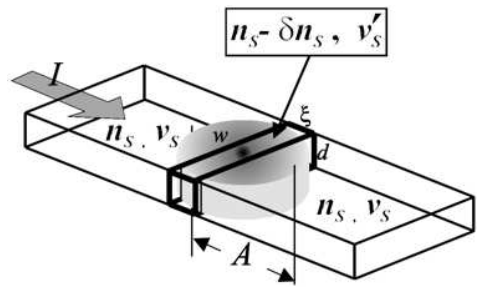}
}
\end{center}
\caption{Schematics of the superconducting strip-line carrying a
current $I$. Bold lines demarcate the smallest volume where a
reduction of the superconducting electron pairs causes a change
of their mean velocity. Grey cylinder depicts the cloud of
nonequilibrium quasiparticles with the size $A$. }
\label{fig:5}       
\end{figure}

Assume a strip made of a superconducting film carries a supercurrent $I<I_C$ as
it is shown in Fig.\ \ref{fig:5}. The film thickness $d$ is chosen smaller
than both the line width $w$ and the coherence length $\xi$
assuring the applicability of a two-dimensional diffusion model
for nonequilibrium electrons. The small film thickness results in
a large magnetic penetration length that exceeds the strip width
even at temperatures well below the superconducting transition
temperature. In this geometry, the magnetic field of the
supercurrent may be neglected and, in the absence of magnetic
vortices, the local supercurrent density should be constant
throughout the cross-section of the strip. However, since our
strips are wider than the coherence length, one can not exclude
the presence of vortices. If there are vortices they should obey a
Kosterlitz--Thouless (KT) topological transition. Square resistance
of our NbN films was always smaller than the resistance quantum
$h/e^2$ resulting in the KT transition temperature \cite{Mooij1}
only slightly below $T_C$. At typical operation temperatures
$\approx T_C/2$ all vortices are then bundled into pairs and do
not contribute to the resistance unless the bias current is very
close to the critical current. Although bundled vortices modulate
the supercurrent density, we will neglect this current
distribution for a while and associate the local supercurrent
density $j = e\, n_sv_s$ with the mean velocity $v_s$ and the mean
density $n_s$ of paired electrons. The local current state may
noticeably change only if $n_s$ changes over a distance $\xi$ or
larger along the current path, shorter perturbations are tunnelled
by electron pairs without energy dissipation. The smallest volume
of the strip relevant to a current change, the $\xi$-slab, is
marked in Fig.\ \ref{fig:5} with bold lines. If $n_s$ in the $\xi$-slab
decreases by an amount $\delta n_s$, the mean pair velocity
increases and becomes
\begin{equation}\label{eq1}
  v_s^\prime = \frac{n_s}{n_s-\delta n_s}\:v_s
\end{equation}
as required by charge flow conservation. The characteristic
conversion time of the pair velocity practically equals the
Ginzburg--Landau relaxation time $\approx h/\Delta$ ($\Delta$ is
the temperature dependent energy gap) and is small compared to the
electron thermalization time $\tau_{th}$. Thus, the mean velocity
in the slab instantaneously follows changes of the pair density.
The $\xi$-slab switches into the normal state if the pair velocity
exceeds the critical value $v_{sc}$ that corresponds to the
critical current density $j_C = e\,n_sv_{sc}$ in the absence of
perturbations.

The concentration of nonequilibrium electrons $C(r,t)$ at a
distance $r$ from the photon absorption site evolves in time due
to multiplication of electrons and their diffusion
\begin{equation}\label{eq2}
C(r,t) = \frac{M(t)}{4\,\pi D\,d\,t}exp(-\frac{r^2}{4\,D\,t}),
\end{equation}
where $M(t)$ is the time dependent number of nonequilibrium
electrons. At $t = \tau_{th}$ nonequilibrium electrons have
thermalized to the energy level $\Delta$ effectively becoming
quasiparticles. Their number reaches the maximum value
$M(\tau_{th}) = \zeta\,\epsilon /\Delta$ where $\zeta\leq 1$ is
the efficiency of the QP multiplication. After the thermalization
time $\tau_{th}$ has elapsed and before subgap phonons appear, the
concentration of nonequilibrium quasiparticles locally equals the
reduction of superconducting electrons $\delta n_s$. Assuming that
well below the transition temperature $n_s\approx N_0\Delta$, one
can rewrite (\ref{eq1}) and find the smallest number of nonequilibrium QP
that is sufficient for switching the $\xi$-slab into the normal
state
\begin{equation}\label{eq3}
    \delta N^* = N_0\Delta\,\xi\,w\,d(1-I/I_C).
\end{equation}
If the spread of the QP cloud is comparable to the coherence
length, all quasiparticles contribute to the change of the
condensate velocity and $\delta n_s = M(\tau_{th})/(w\,d\,\xi)$.
If the cloud is larger (see Fig.\ \ref{fig:5}), only quasiparticles confined
within the $\xi$-slab have to be taken into account. The size
$A(t)$ of the electron cloud is twice the radius of the spot that
confines all nonequilibrium electrons. Equating $M(t)$ to the
integral of the electron concentration (Eq.\ (\ref{eq2})) over the cylinder
with the radius $A(t)/2$ and the thickness $d$, we find the time
dependent size of the electron cloud $A(t) = 4 [D\, t\,
\ln(M(t))]^{1/2}$. If $M(\tau_{th})\geq 100$ and typical
parameters of NbN films are considered, $A(t)$ exceeds the
coherence length already at an early thermalization stage $t <
\tau_{th}$. For $A\gg \xi$  the absolute number $\delta N$ of
nonequilibrium electrons confined to the $\xi$-slab can be
evaluated analytically. Integrating the electron concentration (Eq.\ (\ref{eq2}))
over the slab results in
\begin{equation}\label{eq4}
    \delta N = M(t)\,\frac{\xi}{\sqrt{\pi\,D\,t}}
\end{equation}
The number of nonequilibrium electrons in the slab peaks at
$t\approx \tau_{th}$ when nonequilibrium electrons can already
be treated as quasiparticles. The cut-off of the quantum
efficiency occurs when the maximum number of nonequilibrium QP in
the slab reaches $\delta N^*$. The criteria becomes $\delta
N(\tau_{th}) = \delta N^*$ that allows to determine the efficiency
of the quasiparticle multiplication
\begin{equation}\label{eq5}
    \zeta = \frac{N_0\,\Delta^2w\,d\sqrt{\pi D\,t}\,(1-I/I_C)}{\epsilon}.
\end{equation}
Using material parameters of our films, we estimated an efficiency
of $\approx 0.1$ in the near infrared range.

The maximum size of the actual normal spot $A_n$ can be estimated
solving the equation $C(A_n/2,\tau_{th}) = N_0\Delta$ and using the
above value for the multiplication efficiency. In our films it turns out that $A_n <
\xi$ and according to the hot-spot model the
formation of a resistive state should not be triggered. When applied to the conventional
films with a stoichiometric composition
\cite{Goltsman1,Verevkin1} this analysis even suggests that no normal
spot at all is formed. Finally, it is worth noting the meaning of the size of the
normal spot that is defined by the criteria $A_n/w = (1-I/I_C)$: it is the diameter of a cylinder that would
confine all nonequilibrium quasiparticles at $t = \tau_{th}$ if
their concentration would have been uniform and equal $N_0\Delta$.

\begin{figure}
\begin{center}
\resizebox{0.38\textwidth}{!}{%
  \includegraphics{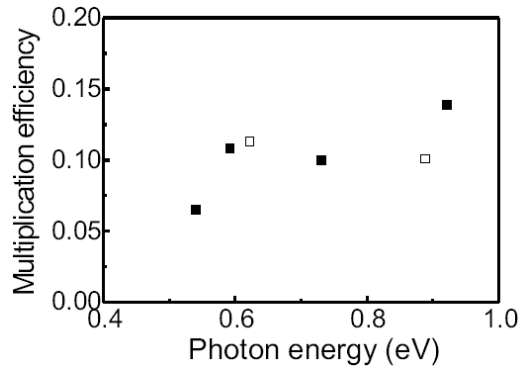}
}
\end{center}
\vspace{0.5cm}       
\caption{Multiplication efficiency of quasiparticles for different
photon energies concluded from current (open symbols) and spectral
(closed symbols) dependence of the photon count rate.}
\label{fig:6}       
\end{figure}

Assuming that the values of the bias currents and wavelengths
indicated by arrows in Figs.\ \ref{fig:3} and \ref{fig:4} define the cut-off that satisfies the criteria
$\delta N = \delta N^*$, we calculated for each pair of current and wavelength the efficiency of QP multiplication.
Fig.\ \ref{fig:6} shows the efficiency $\zeta$ for different photon energies.
The data points presented by open symbols were obtained from the
current dependence of the photon count rate (Fig.\ \ref{fig:4}). They are
less accurate (see the discussion above) than the data obtained
from the spectral dependence of the photon count rate. Therefore,
we omitted the data point derived from the count rate of 0.8~$\mu$m
photons. The efficiency of the QP multiplication increases with
the photon energy. Although the physical reason for the increase is
not quite clear, we speculate that it might be due to a splitting
of the electron cascade into the electron and phonon branches.
The splitting occurs \cite{Kozorezov1} around the Debye energy,
which is slightly less than photon energies used in our
experiment.

\section{Fluctuations and quantum efficiency beyond the cut-off}
\label{sec:5} Besides the fundamental interest in mechanisms,
which result in dark counts in two-dimensional films, fluctuations
are important for sensor applications where small photon fluxes
have to be detected. Depending on the cross-section dark counts in one-dimensional superconductors
may originate from the
spontaneous formation of either classical \cite{Tinkham1} or
quantum \cite{Lau1} phase-slips. When the cross-section confining
the current path extends over the coherence length, dark counts
can appear due to discrete fluctuations of the superconducting
order parameter \cite{Garz1,Engel1} and/or number fluctuations in
the gas of bundled magnetic vortices \cite{Voss1}. It has been
observed in many experiments that the dark count rate strongly
depends on the bias current. Since at temperatures well below $T_C$ the
density of the superconducting condensate does not vary much with
the bias current \cite{Anthore1}, we suggest that the current
dependence of the dark count rate is connected only to the
velocity of the superconducting electrons. In particular, the
velocity controls the minimal excitation energy of the electron
pairs. Although variations of the pair density modify the current
dependence of the dark count rate, experimental data show
\cite{Engel1} that this modification is a week factor in
comparison to the current dependence itself. According to Eq.\ (\ref{eq1}),
the bias current corresponding to the spectral cut-off at a fixed
photon energy can be presented as $I^* = I_C (1-\delta n_s/n_s)$.
For currents $I < I^*$ a photon counting event can be
thought of as a dark count event associated with that portion of
the superconducting strip, where the density of electron pairs has
been decreased due to photon absorption. The elevated pair velocity
in this portion of the strip is determined by the relative
decrease of the mean pair density $\delta n_s/n_s$. In the absence
of perturbations, pairs would have the same velocity at a larger
current $I_1 = I (1-\delta n_s/n_s)$. Therefore, we suggest that
the photon count rate at $I < I^*$ equals the dark count rate at
the current $I_1 = I\,(I_C/I^*)$.

Thin solid lines in Fig.\ \ref{fig:4} simulate the photon count rates beyond the
cut-off. They were obtained by scaling down the current dependence
of the dark count rate (dotted line in Fig.\ \ref{fig:4}). The scaling factor
was used as the only fitting parameter. Visually, the scaling factor
appears as the normalized current corresponding to the cross-point
of the fitting line and the level of the photon count rate at $I =
I_C$ (straight horizontal line in Fig.\ \ref{fig:4}). These values
are marked with arrows and were used for the estimate of the
QP multiplication efficiency, excluding the one
for a wavelength of 0.8~$\mu$m. The method holds as long as the number of
electron pairs destroyed by the photon in the $\xi$-slab is not
very large or, equivalently, the number of remaining pairs
compares to their equilibrium number. If the pair density
changes noticeably, the current dependence of the fluctuation
probability will change like it does with temperature. We
believe that breaking of this condition did not allow us to
satisfactorily fit the photon count rate for the most energetic quanta.
An obvious spin-off of this approach is an understanding of the
spectral dependence of the photon count rate beyond the cut-off.
The spectral dependence should repeat the temperature dependence
of the dark count rate at a fixed bias current. Available
experimental data \cite{Goltsman1,Engel1} qualitatively confirm
this conclusion.

\section{Energy resolving capability}
\label{sec:6} The normal conducting domain created in the meander
by the joint action of a photon and the bias current relaxes via
thermal phonons, which pass the excess energy to the substrate.
Since the relaxation rate is fixed, the life-time of the resistive
state in the meander and, correspondingly, the duration of the
voltage pulse increases with the full excess energy deposited in the film by
the perturbation. Thus the duration of the voltage pulse should
correlate with the energy of the photon, which initiated the
formation of the resistive state. In this section we show that the
resistive state indeed memorizes the photon energy.

\begin{figure}
\begin{center}
\resizebox{0.35\textwidth}{!}{%
  \includegraphics{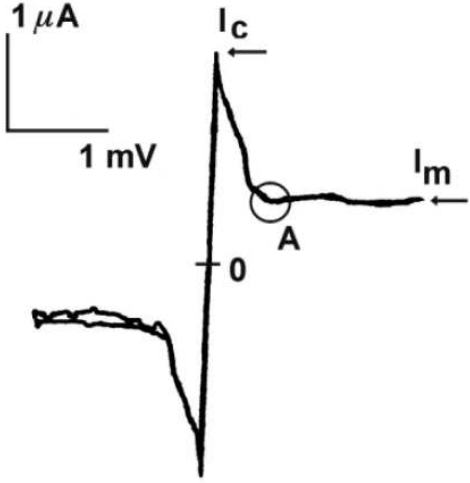}
}
\end{center}
\vspace{-0.2cm}       
\caption{Current--voltage characteristic of the NbN meander at 2 K.
The critical current $I_C$ and the current $I_m$ equilibrating the
walls of the normal domain are both marked by arrows. The area A
(marked by the circle) corresponds to the smallest stable normal
domain of a size approximately equal to the thermal healing
length.}
\label{fig:7}       
\end{figure}

A current--voltage curve of the NbN meander recorded in the voltage-bias
regime at 2~K is shown in Fig.\ \ref{fig:7}. After the critical current value
is reached, the current drops signaling the formation of a normal
domain. The point A corresponds to the smallest stable domain,
which has a length of the order of the thermal healing length $L_T
= (D\,\tau_E)^{1/2}$ where $\tau_E$ is the effective electron
cooling time. When the voltage increases further, a plateau occurs
at an almost constant current $I_m$ that maintains the domain in
equilibrium. According to Ref.\ \cite{Gurevich1}, exactly at this current in a
current biased meander the velocity of the domain walls would
equal zero. The Stekly parameter, which is a measure of the
self-heating by the bias current, can be evaluated as
$\alpha\approx (I_C/I_m)^2 = 25$. Using the resistivity of our
specimen and the dc resistance $\approx 1.2$~k$\mathrm{\Omega}$
corresponding to point A, we estimate a thermal healing length
$L_T \approx 60$~nm. With the diffusivity of $3.5\cdot10^{-5}$~m~sec$^{-1}$ this length implies an effective electron cooling time
$\tau_E \approx 100$~ps. Using this electron
cooling time, the measured critical current density $5.5\cdot10^9$~A~m$^{-2}$
and the electron specific heat $c_E \approx 2\cdot10^2$~J~m$^{-3}$K$^{-1}$, we independently estimated the value of the
Stekly parameter as it follows from the dynamic theory
\cite{Gurevich1} of an electro-thermal domain $\alpha = \rho
j_{C2} \tau_E/c_E(T_C-T)$. We found $\alpha \approx 34$ in fair
agreement with the value concluded from the $I$--$V$ curve. The
steady-state parameters deliver a self-consistent description of
the electro-thermal domain in our specimen ensuring reliability of
the following consideration.

The Stekly parameter $\alpha\,\gg\,1$ implies that even when
$I\,\ll\,I_C$ self-heating strongly influences the dynamics of the
normal domain. The critical energy that initiates the thermal
roll-off or quenching in a current-biased meander can be presented
\cite{Gurevich1} as
\begin{eqnarray}\label{eq6}
    Q_0 & = &
              L_T\,w\,d\,c_E\,(T_C-T)\,\alpha\,i^2\ \ln(\frac{\alpha\,i^2}{\alpha\,i^2-2\,\theta}),\\
        &   & \mathrm{with}\ \theta = \frac{T_C(I)-T}{T_C-T}\ \mathrm{and}\ i = \frac{I}{I_C},\nonumber
\end{eqnarray}
where $T_C(I)$ denotes the temperature, for which the critical current $I_C(T)$ equalizes
the bias current $I$. Assuming the standard
temperature dependence of the critical current, we find for the NbN
meander at $I/I_C = 0.8$ a critical energy $Q_0 = 0.51$~eV that
is $\approx 3$ times smaller than the energy of a 0.8--$\mu$m
photon. Depending on the time constant of the bias and on the
strength of the negative feedback, the quenching may partly occur
also in the voltage-bias regime. To fully avoid quenching the time
constant of the bias should be noticeably smaller than the
electron cooling time. The rates of photon and dark counts at
different biases are shown in Fig.\ \ref{fig:4}. The highest count rate of
$\approx 5\cdot10^7$~sec$^{-1}$ was measured at $I \approx
0.98\,I_C$. Such count rate corresponds to a recovery time of
$\approx 20$~ns that coincides with the full duration of the
signal transient seen in Fig.\ \ref{fig:1}. We believe that the recovery time
more than 10 times larger than the bias time constant $\tau_B \approx 0.5$~ns was most likely
determined by the lower edge of the amplifier band-pass.

\begin{figure}
\begin{center}
\resizebox{0.38\textwidth}{!}{%
  \includegraphics{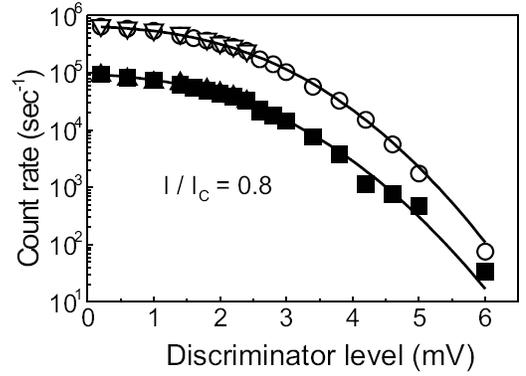}
}
\end{center}
\vspace{0.2cm}       
\caption{The rate of dark (closed symbols) and photon (open
symbols) counts at different discriminator levels. Solid lines are
best fits assuming normal statistical distributions of the pulse
amplitudes.}
\label{fig:8}       
\end{figure}
\begin{figure}
\vspace{0.2cm}
\begin{center}
\resizebox{0.32\textwidth}{!}{%
  \includegraphics{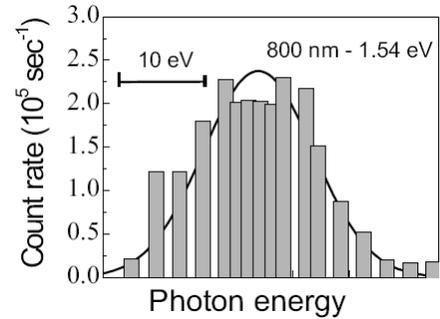}
}
\end{center}
\vspace{0.2cm}       
\caption{Spectrum of 0.8--$\mu$m photons recorded with the NbN
meander at 2~K.}
\label{fig:9}       
\end{figure}

At large bias currents very close to $I_C$ the noise background may be
so high that dark counts outnumber photon counts. Accordingly, uncertainties
in the photon count rate increase. Therefore, we analyzed the statistical
distribution of the voltage transients at a smaller bias $I =
0.8\,I_C$ when the rate of dark counts was at least an order of
magnitude less than the rate of photon counts. Fig.\ \ref{fig:8} shows the
dark count rate and the photon count rate for 0.8--$\mu$m photons
measured at different discriminator levels of the pulse counter.
When the discriminator level is set low all events are counted.
Increase of the discriminator level leaves more and more events
uncounted until at a high enough level none is counted anymore.
Assuming a normal distribution of the signal amplitude, we fit
the integral of the distribution to the experimental data.
Best-fit curves are shown in Fig.\ \ref{fig:8}. Two fitting parameters were
used: the mean value and the dispersion of the normal
distribution. We have found a best dispersion value
$1.65\,\pm\,0.03$~mV for both dark and photon counts and mean
values 1.7~mV and 1.9~mV for dark and photon counts, respectively.
In terms of the strength and statistical distribution, dark counts
are identical to photon counts when photons have the critical
energy $Q_0$.The wavelength of a photon having this critical
energy is 2.4~$\mu$m perfectly matching the longest
wavelength we detected with this meander. Relating the mean values
of the signal amplitudes to the energies of 0.8--$\mu$m and
2.4--$\mu$m photons we found a scaling factor of 0.14~mV/eV. Fig.\ \ref{fig:9}
presents the spectral histogram for 0.8--$\mu$m photons retrieved
with this scaling factor from the experimental data. The best-fit
envelope shown by the solid line represent a normal distribution
with a dispersion of 6.5~eV. Although the effect has an energy
resolving power, it is far too low for practical purposes. Factors
limiting the energy resolution are discussed below.

We should first estimate the energy dissipated in the meander by
the bias current. After a photon has released an energy larger
than $Q_0$, the normal domain starts to grow. The propagation
velocity of the domain walls, i.e.\ the superconductor-normal
interfaces, determines the growth rate of the normal domain. The
velocity decreases as the actual bias current drops following the
domain growth. In our experiment the life-time of the domain
roughly coincides with the time constant $\tau_B$ of the bias
circuit. The size of the domain, and consequently the energy
dissipated by the current, can be determined integrating the Joule
power with the particular current dependence of the interface
velocity. Considering the simple step-edge heat-dissipation model,
it can be shown \cite{Gurevich1} that the interface velocity v
increases linearly with the current:
\begin{equation}\label{eq7}
    v = (\alpha\,\frac{D}{\tau_E})^{1/2}\ (i-i_m)\ \theta^{-1/2},
\end{equation}
where $i_m = I_m /I_C$. With this dependence the expression for
the Joule energy dissipated within the time $\tau_B$ takes the
form
\begin{equation}\label{eq8}
    E_B =
    \frac{1}{5}\,I^2\,\rho\,(\frac{D}{\tau_E})^{1/2}\ \frac{\tau_B^2}{w\,d}.
\end{equation}
Using $\tau_B \approx 0.5$~ns we found a current contribution of
$\approx 11$~eV that is almost an order of magnitude larger than
the energy of a 0.8--$\mu$m wavelength photon. Given the linear
dependence of the response duration on the total energy released
in the meander, the ratio of the mean amplitudes of photon and
dark counts (see Fig.\ \ref{fig:8}) correlates well with the energy
contributed by the bias. Denoting by $\epsilon_1$ and $\epsilon_2$
the energies of 0.8--$\mu$m and 2.4--$\mu$m photon, respectively, we
find $(E_B+\epsilon_1)/(E_B+\epsilon_2) \approx 1.09$ in
agreement with the amplitude ratio 1.1 concluded from the
experiment.

The dispersion of the life-time of the normal domain may originate
from structural defects of the meander mediated by the domain
length. In NbN films defects are most likely boundaries between
grains, which have a typical size of ten nanometers
\cite{Semenov3}. Intuitively, the dispersion has a maximum value
when the domain length is of the order of the grain size.
Contribution of the film structure to the measured dispersion
seems plausible because the normal domain in our meanders covers
only a few grains. The cooled amplifier also contributed to the
measured dispersion since its equivalent voltage noise was only
slightly less than the amplitude of the voltage pulse delivered by
the meander. With a SQUID amplifier that typically has a subkelvin
noise temperature, the energy resolution should drastically
improve. Finally, using a superconductor with a smaller energy gap
one should be able to increase the conversion ratio of the photon
energy to the domain length. Given the same dispersion this will
also lead to a better energy resolution. An estimate shows that
the cumulative effect of all factors mentioned should suffice for
an energy resolution comparable to the resolution of conventional
superconductor photon detectors.

In summary, we observed the current dependent spectral cut-off in
the appearance of the resistive state after absorption of a single
near infrared photon in nano-structured NbN superconducting films.
We generalized the normal-spot model explicitly taking into
account the diffusion profile of nonequilibrium quasiparticles.
With this approach, we found a super-linear photon energy
dependence of the quantum yield in NbN that was attributed to a
splitting of the thermalization process into the phonon and the
electron cascades around the Debye energy. The resistive state was
shown to memorize the energy of the photon, which had initiated
it. Ultimate energy resolving capability of this effect should be
comparable with the energy resolution of conventional detectors.


\begin{thebibliography}{}
\bibitem{Rem1}
Direct measurements of the magnetic penetration length indicated
normal conducting $\approx\,0.4$ nm thick layers at the film
interfaces with the substrate and air and also $\approx\,4$ nm
wide normal conducting areas along the stripe edges. Together with
the observed 15\% non-homogeneity of the stripe width this reduces
the electrical cross-section of our meanders to 0.58 of the
geometrical cross-section.
\end{thebibliography}
\end{document}